\documentstyle[twoside,fleqn,espcrc2,epsfig]{article}

\newcommand{\AmS}{{\protect\the\textfont2
  A\kern-.1667em\lower.5ex\hbox{M}\kern-.125emS}}

\title{Atmospheric Neutrino Flux: A Review of Calculations}

\author{T.K. Gaisser\address{Bartol Research Institute,
        University of Delaware,\\
        Newark, DE 19716, USA}%
        \thanks{Research supported in part by the U.S. Department
            of Energy under Grant No. DE-FG02-91ER40626}}

\begin{document}
\begin{abstract}
Interpretation and understanding of the evidence for neutrino
oscillations depends on knowledge of the atmospheric neutrino
beam.  In this talk I assess how well various features are known.
The goal is to determine to what extent uncertainties in the neutrino beam
may limit the conclusions about neutrino properties and
which features of the evidence for neutrino oscillations
are most robust.
\end{abstract}

\maketitle

\section{Introduction}

\begin{figure}[htb]
\flushleft{\epsfig{figure=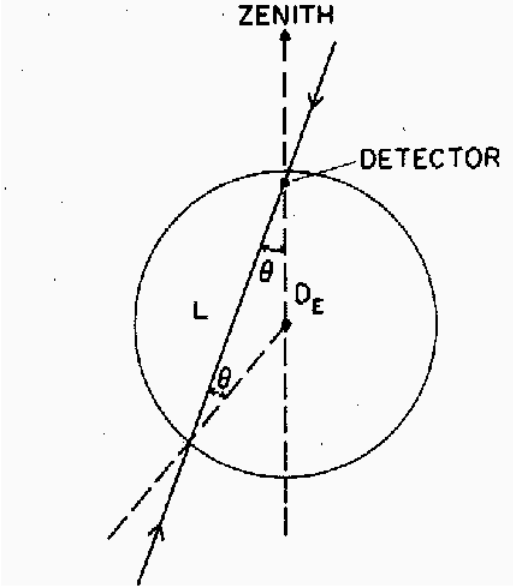,width=7cm}}
\caption{Illustration of up-down symmetry of atmospheric
neutrinos in the absence of effects of the geomagnetic
field and assuming no oscillations.}
\label{fig:geometry}
\end{figure}

The concept of using the atmospheric neutrino beam to look for
neutrino oscillations is illustrated in Fig. 1~\cite{Ayresetal}.
With a single detector it is possible simultaneously to cover
a range of pathlengths from $\sim10$ to $\sim10^4$~km, corresponding
respectively to downward moving and upward moving neutrinos.
The atmospheric neutrino beam has an energy spectrum determined
by the steeply falling primary cosmic-ray spectrum, which generates
the neutrinos by interactions
of the cosmic ray nucleons in the atmosphere.
Examples of calculated neutrino fluxes are shown in Fig. 2.

\noindent
\begin{figure}[htb]
\flushleft{\epsfig{figure=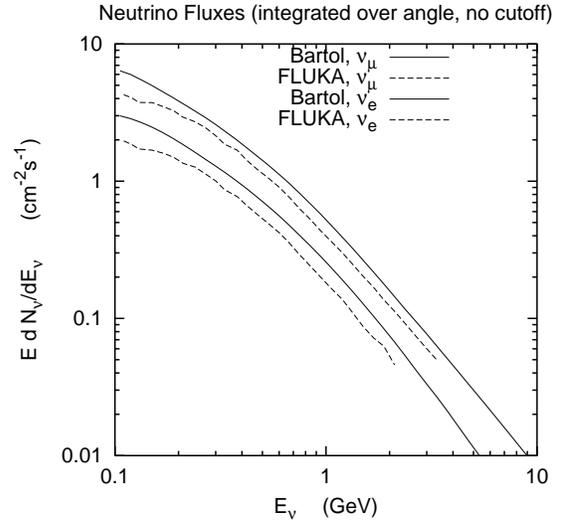,width=7.5cm}}
\caption{Fluxes of atmospheric neutrinos from two calculations:
FLUKA~\cite{Giuseppe3D}, dashed; Bartol~\cite{AGLS}, solid.}
\label{fig:nufluxes}
\end{figure}

The neutrino spectrum, which falls with energy,
must be folded with the rising neutrino cross section
to obtain the expected event rate as a function of neutrino energy.
Most neutrino interactions are in the range from a few hundred MeV
to a few tens of GeV.  Thus atmospheric neutrinos have the
potential to probe the range $1 < L/E < 3\times 10^4$~km/GeV.
From the standard two-flavor oscillation equation,
\begin{equation}
P_{\nu_\mu\nu_\mu}\,=\,1\,-\,\sin^22\theta\sin^2\left[ 1.27 \delta m^2
(eV^2){L_{km}\over E_{GeV}}\right],
\label{oscillation}
\end{equation}
we find, therefore, that the atmospheric neutrino beam can in principle
probe
down to $\delta m^2$ as small as $\sim 2\times 10^{-5}$~eV$^2$.
(The actual lower limit to sensitivity in $\delta m^2$
is somewhat higher than this because of the relatively large
scattering angle between neutrino and lepton in charged
current interactions of low energy neutrinos.)

Atmospheric neutrinos originate with the $\pi\rightarrow\mu\rightarrow e$
decay chain.  Therefore at sufficiently low energy such that
muons as well as pions decay, one expects
\begin{equation}
{\nu_e +\bar{\nu}_e\over \nu_\mu +\bar{\nu}_\mu}\,\approx\,{1\over 2}.
\end{equation}
The anomalous value of this ratio\cite{IMBGeV,KamGeV,Soudan,SuperK} 
(for which many sources of uncertainty
cancel in the calculations) suggests neutrino oscillations
involving $\nu_\mu$ and/or $\nu_e$ as a possible explanation.
The telltale evidence~\cite{evidence} for oscillations as the source of this
anomaly comes from the pathlength dependence of the neutrino fluxes.
The energy-dependence of the up-down asymmetry for muon-like
events, coupled with the
fact that events initiated by electron neutrinos appear to
have the expected energy and angular dependence, indicates that
the primary effect is oscillations involving muon neutrinos
with large mixing 
and $\delta m^2\approx 3\times 10^{-3}$~eV$^2$~\cite{evidence}.

In the remainder of this paper I will describe how the geomagnetic
field affects the interpretation of the atmospheric neutrino data,
review the primary cosmic-ray spectrum, discuss the uncertainties
in the treatment of pion production and mention the comparison
of the calculations to measurements of muons high in the atmosphere.
It is important to review these points at this time because of
the recent publication of two new calculations~\cite{Giuseppe3D,Waltham},
which are three-dimensional, as compared to previous one-dimensional
calculations.

\section{Geomagnetic field effects}

In the absence of oscillations, the only significant deviation
from isotropy of the atmospheric neutrino beam is caused by
the geomagnetic field, which prevents low energy primary cosmic
rays from reaching the atmosphere at low geomagnetic latitudes.
For example, at Kamioka the geomagnetic cutoff is $\approx 10$~GeV
for primary protons near the vertical.  The downward neutrino
flux at Kamioka is therefore lower than, for example, at Soudan,
where the vertical cutoff is negligible.  Approximately half
the downward neutrinos with $E_\nu < 1$~GeV at Soudan
come from primary protons with $E<10$~GeV.  This contribution is
absent from the downward neutrino flux at Kamiokande.  As a
consequence the downward flux of sub-GeV neutrinos at Kamiokande
is roughly half that at Soudan.  The neutrino flux from the lower hemisphere
is similar at the two detectors and intermediate between the
downward fluxes at the two locations.  This is because the neutrinos
from below are produced over a large fraction of the Earth's surface
and so average over a range of high and low geomagnetic cutoffs.
Thus the up-down asymmetry
at each detector location is a combination of geomagnetic effects
together with the effects of any oscillations that may be
present, and the combination depends on detector location.
Moreover, the geomagnetic effects become less important as energy
increases.  It is therefore useful to start by considering the 
atmospheric neutrino analysis without the geomagnetic field.

For a flux of primary cosmic rays that is spatially isotropic,
the atmosphere is a spherical shell source of neutrinos
with equal luminosity per unit volume independent of latitude
and longitude.  The number of neutrinos produced per unit
volume of atmosphere into solid angle $d\Omega$,
\begin{equation}
S(\theta,h)\;=\;{dN_\nu\over d\Omega\,dV},
\end{equation}
depends only on local zenith angle, $\theta$, and altitude, $h$.
In these circumstances one can show from the geometrical
construction of Fig. 1, that the neutrino flux is
up-down symmetric; i.e. symmetric about
$\cos\theta\leftrightarrow -\cos\theta$.
Since neutrinos from decay of pions, muons and other secondary cosmic
rays are produced over a range of altitudes peaking around
15 to 20~km above sea level~\cite{pathlength}, it follows that
local variations in surface altitude introduce
a negligible  deviation from this symmetry.
The symmetry also requires that differences
caused by local variations of pressure are negligible.

A detector like Super-Kamiokande has an acceptance that
is up-down symmetric.  In the absence of geomagnetic effects, therefore,
a simple measurement of the up-down asymmetry
as a function of energy is a probe of neutrinos oscillations.
Therefore the simplest and most
robust evidence for oscillations is a deviation from up-down
symmetry in an energy range high enough so geomagnetic effects
are small.  Fig. 3~\cite{EGS} shows the distribution of neutrino energy
for four classes of neutrino interactions.  The corresponding distributions
in energy per nucleon for the primary cosmic-rays are about a factor
of ten higher.  Thus the median primary of the multi-GeV events
is about 50 GeV/nucleon, which is high enough so that geomagnetic
effects are unimportant.  Neutrino-induced upward muons provide
a higher energy sample, but it is not possible to make a simple
up-down comparison because the downward muon flux is dominated
by muons produced from pion decay in the atmosphere rather than
by interactions of neutrinos.
  
\noindent
\begin{figure}[htb]
\vspace{10pt}
\flushleft{\epsfig{figure=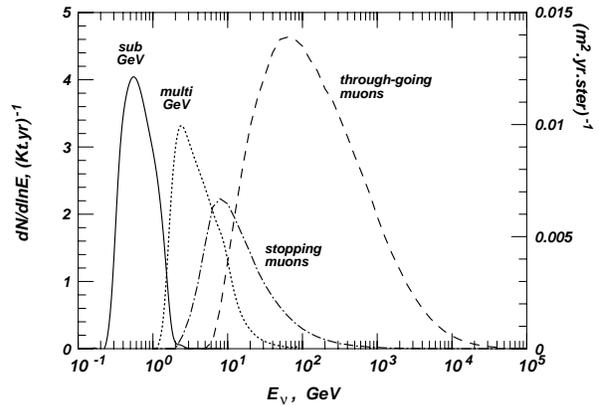,width=8.5cm}}
\caption{Distribution of neutrino energies for various classes
of atmospheric neutrino events.}
\label{fig:response}
\end{figure}

Interpreting the up-down asymmetry of the neutrino fluxes
in the GeV range and below requires that the geomagnetic effects
be well understood.  Geomagnetic effects on the primary cosmic
radiation are indeed very well understood.  They form the
basis for an entire subfield of cosmic-ray physics~\cite{Hillas}.
Low energy particles (few
GeV) at low geomagnetic latitudes cannot reach the atmosphere to interact.
Particles of intermediate energy ($\sim 10$ to $20$~GeV) that reach
the atmosphere show a strong east-west asymmetry, while high
energy particles ($\sim 100$~GeV and higher) are essentially unaffected
by the geomagnetic field.  

The east-west effect is a consequence of bending of positive primaries
in the geomagnetic field.  In fact, it was the observation
of the excess of primary cosmic rays from
the west~\cite{Johnson,AC} from which it could be
deduced that the primaries were predominantly positively
charged.  The expected azimuthal dependence of the neutrino
flux associated with the east-west effect is the same whether
or not the neutrinos oscillate because the distribution of
pathlengths that contribute to a particular zenith angle
band is independent of azimuth.  Comparison between expected
and observed azimuthal distributions is therefore a good
check of the systematics of the whole chain of data analysis and
calculations~\cite{LSG}.  The measurements show the
expected azimuthal dependence for both muon and electron
neutrinos~\cite{eastwest}, indicating that the geomagnetic
effects are well understood.

\section{Primary spectrum}

A standard procedure for calculating the flux of atmospheric neutrinos
is to generate atmospheric cascades for a spectrum of primary protons
and nuclei and form the neutrino spectra in the absence of the
geomagnetic fields.  The resulting neutrino spectra can be filtered through
the geomagnetic configuration relevant for a particular detector
location, discarding those neutrinos produced by primaries that would not
have reached the atmosphere.  An alternative approach~\cite{Perkins}
starts from the measured muon fluxes high in the atmosphere and
uses the genetic relation between neutrinos and muons to obtain
the muon spectrum.  In either case, the normalization of the
neutrino flux depends on the absolute normalization of a measured spectrum 
of charged particles (either the primary cosmic rays or the muons).
The advantage of starting with the muons is that one bypasses
uncertainties in knowledge of pion production, which have
a similar effect on both neutrinos and muons.  A disadvantage
is that the acceptance for muons is sensitive to details of propagation
in the geomagnetic field.  In addition,
the measurements of the primary flux are made with relatively long
exposures at or above the atmosphere while the relevant muon measurements
are generally made during a short balloon ascent.

The mixture of nuclei in the primary spectrum is such that
approximately 80\% of nucleons in the cosmic radiation are free
protons, 15\% are bound in alpha particles and the remainder
are in heavier nuclei.  Spectra of protons and helium with
$E<100$~GeV/nucleon have been measured at the top of the atmosphere
in a series of balloon-borne spectrometer experiments.  Recent
measurements~\cite{CAPRICE,MASS,AMS,IMAX,BESS}
cluster around a lower normalization~\cite{LEAP}
than an earlier standard reference~\cite{Webber}.  At higher energy,
measurements have so far been possible only with calorimeters,
which, because of punchthrough, may have larger systematic uncertainties.
The data are summarized in Fig. 4.
\begin{figure}[htb]
\flushleft{\epsfig{figure=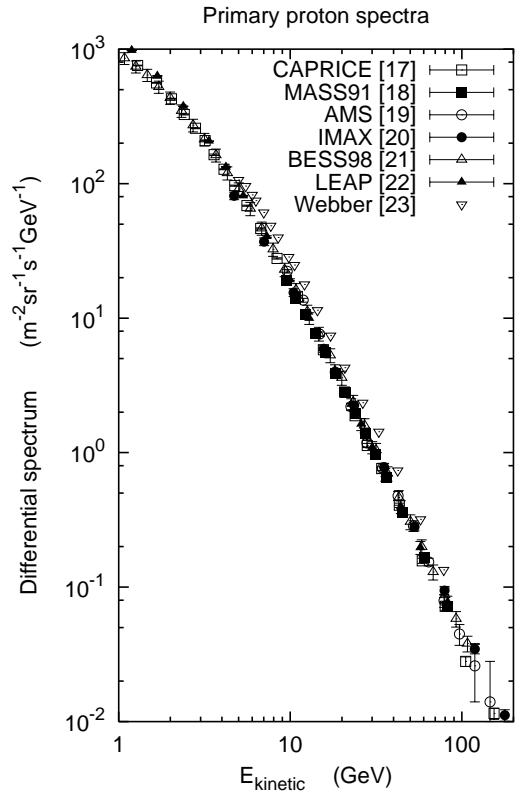,width=7.5cm}}
\caption{Primary spectra.}
\label{fig:spectra}
\end{figure}

\section{Comparison of calculations}

There are now five independent calculations of the neutrino
spectrum that start from the primary cosmic-ray spectrum filtered
by the geomagnetic field.  The calculations of 
Refs.~\cite{AGLS,HKKM,BN} are one-dimensional, assigning
all produced neutrinos the directions of the primary particle
that produced them.  The ingredients of these calculations
have been compared previously~\cite{GHKLMNS}.  Recently 
two three-dimensional calculations 
have been published~\cite{Giuseppe3D,Waltham}.  Treatment
of the geomagnetic cutoffs, which now depend on 4 variables
instead of 2, makes the calculation significantly more complex.
These calculations are very important because they
check the major, technical simplifying assumption
made by the previous calculations.

A conclusion reported in Ref.~\cite{Giuseppe3D} is that
differences for
predicted event rates and how they depend on direction
and energy are relatively small between one-dimensional and
three dimensional versions of the same calculation.
This has the important consequence that the simpler
one-dimensional calculations can be used to explore
the consequences of uncertainties in input to the calculations.
Fig. 2 compares the Bartol neutrino flux~\cite{AGLS} with
the 3-dimensional flux of Ref.~\cite{Giuseppe3D}.  Both calculations
use the primary spectrum of Ref.~\cite{AGLS}, and the
geomagnetic field has been turned off.  Most of the difference,
therefore, is presumed to be
caused by differences in the treatment of pion production.

\section{Pion production}
Most pions with $E< 100$~GeV in the atmosphere decay before
they interact.  Therefore, for neutrinos in the sub- and multi-GeV
range, only interactions of protons and helium play a significant
role.  The most important information needed about these interactions
is the inclusive cross sections for pion production.
The important range of interaction energies extends up to $\sim 100$~GeV
for sub-GeV and $\sim 1$~TeV for multi-GeV interactions.  The most
probable energy of a primary proton for a sub-GeV event at
Super-K is $\approx 20$~GeV. The corresponding number at Soudan
is about $10$~GeV because of the lower geomagnetic cutoff.  The
parent energies for multi-GeV events are correspondingly higher.
The distributions of secondary nucleons is also important because
a significant fraction of the neutrino production occurs in secondary
or tertiary interactions of the nucleons.  Neutral pions (and $\eta$ mesons)
are also important in the sense that energy deposited in
the electromagnetic part of the cascade is not available
for production of neutrinos.  Distributions of kaons
begin to be important for multi-GeV events (and they are dominant
for neutrino-induced upward muons~\cite{AGLS}).

Data on pion production have been discussed recently in Ref.~\cite{EGS}.
Existing measurements cover a significant fraction, but not all,
of the relevant range of phase space for charged pion production
in proton collisions on beryllium and aluminum.  New, more
precise measurements covering all phase for proton
interactions on a range of light nuclei
(including nitrogen and oxygen) would be of great interest.
Use of a helium beam would also be of interest.

The difference between the neutrino calculations shown in Fig. 2 
most probably mainly reflects a difference in the fraction of energy 
going into production of charged pions.
The calculation of Ref.~\cite{AGLS} uses a phenomenological
hadronic event generator (TARGET) developed for cosmic-ray cascade calculations.
FLUKA~\cite{FLUKA} uses a more sophisticated microscopic
model of particle production with intranuclear cascading.  It
incorporates the event generator as an integral part
of a cascade code capable of simulating interactions and cascades in
complex detector geometries.  The philosophies are quite different.
The FLUKA interaction model is tested and adjusted by
comparing directly to double differential cross sections for
a wide range of data sets {\it as measured}
$$
(e.g.\;\;E_\pi\,{dN_\pi(E_p)\over dp_L\,d^2p_T}\;).
$$
The strategy with TARGET is to fit $p_T$ distributions at
each longitudinal momentum, $p_L$, extrapolate into
unmeasured regions of phase space at each $p_L$, and integrate to obtain
the energy flow into each secondary channel,
$$
E_\pi\,{dN_\pi(E_p)\over dE_\pi}.
$$
A detailed investigation of the sources of difference between
these two models and others~\cite{HKKM}, and their implications,
is currently in progress.

\section{Comparison to atmospheric muons}

Most muons and neutrinos are produced between 10 and 30 kilometer
altitudes~\cite{pathlength} in closely related
processes.  Measurements of muons at these
altitudes therefore in principle provide a check of the neutrino calculations
in which the uncertainties in pion production cancel to the
extent that they are common to both the neutrinos and the muons.
Recently there have been several comprehensive measurements
of muons during ascent of balloon payloads~\cite{MASS,CAPRICEmu,HEATmu},
and some discrepancies with calculations have been noted.  Generally,
the agreement is best at float altitude, where only the first
interaction plays a role.   An important limitation to the use
of muons to normalize the neutrinos, however, is that the muons
are more sensitive to details of the calculation.  For example, since most
muons decay in flight, a change in interaction lengths,
which moves the cascade up or down, can change the muon flux
at a particular altitude without changing the corresponding
neutrino flux, which is an integral over the whole atmosphere.
Since muons follow curved trajectories in the geomagnetic field,
approximating by straight lines can have a similar effect (by
moving the muon decays lower in the atmosphere).  In addition, since positive
and negative muons have opposite curvature~\cite{Johnson2} their
response to cutoffs of the primary cosmic radiation are somewhat
different.  Three-dimensional calculations of
atmospheric muons are in progress.~\cite{SCGB}.

\section{Summary}
\begin{itemize}
\item The observed pathlength dependence of the muon-like
events in Super-Kamiokande~\cite{evidence}  points
to neutrino oscillations as the source of the atmospheric
neutrino anomaly.   
\item Three-dimensional calculations of the atmospheric
neutrino flux remove an important approximation present
in previous calculations.  It appears, however, that
the simpler one-dimensional calculations are adequate
for exploring differences among calculations.
\item Uncertainty in the normalization of the primary
spectrum is now reduced to $\approx \pm15$\%, so the
main remaining source of uncertainty is the representation
of pion production.
\item If the neutrino flux is significantly lower than
calculated in Refs.~\cite{AGLS} and~\cite{HKKM}, then
there is a potential problem accounting for the relatively large
number of electron-like events seen in SuperKamiokande
within a predominantly $\nu_\mu\leftrightarrow\nu_\tau$
oscillation scheme.
\item New measurements of atmospheric muons are being considered
in airplanes and with slow balloon ascents.
\item Uncertainties in expected event rates due to the
imprecise knowledge of neutrino cross sections in the GeV
energy range have not been discussed here, but may be
important.~\cite{sigmanu}
\end{itemize}
\noindent
{\bf Acknowledgments}.  I am grateful 
to Todor Stanev and Ralph Engel for collaboration on this work
and to Todor Stanev for reading the manuscript.  I thank
Giuseppe Battistoni and Alfredo Ferrari for providing
information about the calculation of Ref.~\cite{Giuseppe3D}.
I also thank John Ellis and Alvaro De Rujula for hospitality
during ``Neutrino Summer'' at CERN where I began work on this talk.

\end{document}